\documentclass[aps,pra,twocolumn,showpacs,floatfix]{revtex4}
\usepackage{graphicx}
\usepackage{graphics}
\begin{document}

\title{Stabilizing an atom laser using spatially selective pumping and feedback}

\author{Mattias Johnsson, Simon Haine and Joseph J. Hope}
\affiliation{ARC Centre for Quantum-Atom Optics, Faculty of Science, The Australian National University, Canberra, ACT 0200,
Australia}

\date{\today}

\begin{abstract}
We perform a comprehensive study of stability of a pumped atom laser in the presence of pumping, damping and outcoupling. We also introduce a realistic feedback scheme to improve stability by extracting energy from the condensate and determine its effectiveness. We find that while the feedback scheme is highly efficient in reducing condensate fluctuations, it usually does not alter the stability class of a particular set of pumping, damping and outcoupling parameters. 
\end{abstract}


\maketitle

\section{Introduction}
\label{secintroduction}

The experimental realization of Bose-Einstein condensates (BECs) has provided a testbed for many fundamental issues in interacting quantum systems, as well as providing a general tool for investigating aspects of atomic physics such as the behaviour of weakly interacting alkali gases \cite{dalfovoET1999}. One major application that BECs offer is the possibility of creating an atom laser \cite{ballaghET2000}. Just as the optical laser revolutionised optics by offering spatial and temporal coherence, high spectral density and mode selectivity, the pumped atom laser offers the possibility of doing the same for atomic physics. 

Atom lasers have been achieved by outcoupling atoms from a BEC using some external means to change the state of a subset of the atoms in the condensate from a trapped to an antitrapped state \cite{mewesET1997,blochET1999,hagleyET1999}. This can produce a beam of atoms that exhibits both spatial and temporal coherence \cite{andersonET1998,wiseman1997}. As in optical lasers, a narrow linewidth (i.e. small momentum spread) can be attained by ensuring the outcoupling is weak, although this results in a low beam flux \cite{hope1997}. In order to create a high flux, narrow linewidth atomic beam, there must be competition between a depletable pumping mechanism and damping resulting in gain narrowing \cite{wiseman1997}. 

A multimode analysis of an atom laser is possible using semiclassical techniques such as the Gross-Pitaevskii equation \cite{dalfovoET1999,leggett2001}. Such a model can describe spatial variation of the modes and determine if the laser approaches single-mode operation. It cannot, however, calculate the linewidth of the laser in this limit, since the linewidth of a single-mode laser is governed by the quantum statistics of the mode, so that a full quantum analysis is required \cite{wallsET1994}. This drastically increases the difficulty of the calculation, limiting current efforts to only a few modes.

In order for an atom laser to reach single mode operation it is necessary that it is stable, such that the dynamics of the condensate has a steady-state solution. It has previously been shown that if the pumping of the condensate is spatially uniform, then an atom laser will be unstable if the nonlinear interaction strength between the atoms is below some critical value \cite{Haine2002}. This is because increasing nonlinearity causes greater damping of higher frequency trap modes, leading to the promotion of lower trap modes, and ideally the ground state. Stability can also be improved if a spatially dependent pumping scheme is utilized, where the condensate is preferentially pumped in a narrow range near the center of the trap \cite{Haine2003}. As lower frequency modes have a greater overlap with the central portion of the trap, these modes are more effectively populated than higher frequency modes, again leading to the promotion of low-frequency modes and encouraging stability.

Although interatomic interactions can lead to single mode operation of an atom laser, strong interactions will ultimately limit the linewidth, as they will cause phase diffusion of the lasing mode \cite{wisemanET2001,thomsenET2002}.  An independent method of improving modal stability without using high interaction energies is to use a feedback mechanism whereby energy is removed from the condensate using continuous knowledge of the atomic cloud's dynamics to tune the trap parameters. It has been shown that if one considers an isolated condensate in a trap, with no pumping, loss or outcoupling present, then provided the quantities $\langle x\rangle$, $\langle x^2\rangle$ and $\langle |\psi|^2\rangle$ can be measured, where $\langle q\rangle = \int \psi^{\ast} q\psi \,dx $, then it is almost always possible to extract energy from the condensate, and thus draw it towards the lowest energy ground state \cite{Haine2004}. It is possible, however, that some specific (non-ground) state may exist that provides no error signal to the feedback loop causing the feedback process to have no effect. This occurs when all the moments chosen as error signals are stationary, i.e. do not change with time.

The purpose of this paper is to tie these threads together into a comprehensive analysis of atom laser stability, incorporating loss, outcoupling, spatially dependent pumping, a depletable reservoir and feedback. We will consider stability across a broad range of pumping regimes, and include the effects of the outcoupled beam on the condensate as this modifies some of the stability conclusions in Ref.~\cite{Haine2003}.

\section{Model}

As discussed in the introduction, determining the stability of an atom laser does not require a full quantum analysis. We will therefore use a multimode semiclassical model based on the Gross-Pitaevskii equation. We denote the condensate field by $\psi_t({\bf{x}})$ and the untrapped field by $\psi_u({\bf{x}})$. $\psi_t({\bf{x}})$ forms the lasing mode, and is pumped by an incoherent reservoir of atoms which has a density described by $n({\bf{x}})$, with the coupling between the two given by $\kappa(\bf{x})$. For calculational tractability, we restrict ourselves to a one dimensional condensate, so that the dynamical equations for the fields can be written as
\begin{eqnarray}
i\hbar \frac{d\psi_t}{dt} &=& \left[ -\frac{\hbar^2}{2m}\nabla ^2 + V_t -i\hbar\gamma_t^{(1)}
	+ \left( U_{tt} -i\hbar\gamma_t^{(2)}\right) |\psi_t|^2 \right. \nonumber \\
	 && \left. + \left( U_{tu} -i\gamma_{tu}^{(2)}\right) |\psi_u|^2 + \frac{i \hbar}{2} \kappa(x)n(x)\right]
	\psi_t \nonumber \\
	&& + \kappa_{\mathrm{out}}(x) e^{i k x} \psi_u, \label{eqpsitrapped} \\
i\hbar \frac{d\psi_u}{dt} &=& \left[ -\frac{\hbar^2}{2m}\nabla ^2 + mgx
	+ \left( U_{uu} -i\hbar\gamma_u^{(2)}\right) |\psi_u|^2 \right. \nonumber \\
	&& \left. + \left( U_{tu} -i\gamma_{tu}^{(2)}\right) |\psi_t|^2 \right]\psi_t 
	+ \kappa_{\mathrm{out}}(x) e^{-i k x} \psi_t, \label{eqpsiuntrapped} \\
\frac{dn}{dt} &=& r-\gamma_p n(x) -\kappa(x)|\psi_t|^2 n(x) +\lambda\nabla^2 n(x), \label{eqreservoir}
\end{eqnarray}
where $m$ is the atomic mass, $V_t$ is the trapping potential, $g$ is the acceleration due to gravity (taken to be in the negative $x$ direction), $U_{ij} = 4\pi\hbar^2 a_{ij}/m$ is the interatomic interaction between $\psi_t$ and $\psi_j$ and $a_{ij}$ is the s-wave scattering length between those same fields. $\gamma_i^{(1)}$ is the loss rate of $\psi_i$ due to background gas collisions, $\gamma_i^{(2)}$ is the loss rate of $\psi_i$ due to  two-body inelastic collisions between particles in that state, $\gamma_{tu}^{(2)}$ is the loss rate of each field due to two-body inelastic collisions between particles in the other electronic state, $\kappa_{\mathrm{out}}$ is the coupling rate between the trapped field and the output beam, $k$ is the momentum kick due to the coupling process, $r$ is the rate of density increase of the incoherent cloud of atoms forming the reservoir, $\gamma_p$ is the loss rate of that cloud and $\lambda$ is the spatial diffusion coefficient. We choose the coupling between the reservoir and the trapped field to be of the form
\begin{equation}
\kappa(x) = \kappa_0 e^{-x^2/\sigma^2}
\end{equation}
enabling us to consider a spatially dependent pumping scheme.

The pumping terms in the above equations are phenomenological, describing an irreversible pumping mechanism from a reservoir which can be depleted but is replenished at a steady rate. These two features are necessary for any pumping mechanism that generates gain-narrowing through the competition of the gain and loss processes of the lasing mode.
We have not included three-body losses, which can be important. Near a Feshbach resonance they may be negligible, however, and still allow a wide range of scattering lengths \cite{Roberts2000}.

To model the feedback stabilization scheme we adopt our previous approach \cite{Haine2004}, and assume that we can control the condensate in some fashion in response to a set of continuously measured error signals. Realistic control parameters we can use to affect the condensate will be the position of the trap minimum potential, the trap strength, and the nonlinear interaction strength. Consequently we assume it is possible to control an external potential of the form $V_{fb}^1 = a_1(t) x + a_2(t) x^2$ and a nonlinear interaction strength of the form $V_{fb}^2 = b(t) |\psi_t|^2$. The $a_1(t)$, $a_2(t)$ and $b(t)$ correspond to time-dependent control parameters that can be manipulated according to the measured error signals. Altering $a_1$ and $a_2$ corresponds to changing the position of the trap minimum and its curvature, while tuning $b(t)$ corresponds to manipulating the nonlinear interaction strength between the atoms in the trap. The latter can be accomplished by controlling magnetic field close to a Feshbach resonance \cite{Roberts2001}. This is equivalent to controlling the bias magnetic field in a magnetic trap, or applying a constant magnetic field in an optical trap, and has been achieved with high precision \cite{Donley2001}.
Provided the parameters $a_1$, $a_2$ and $b$ are chosen correctly, the rate of change of energy of the condensate must always be non-positive. This conclusion assumes that the condensate is not pumped and has no losses. As we shall see later, the presence of such features can mean that this feedback scheme is not always guaranteed to remove energy from the condensate, although in practice it normally still does a very good job.

For our error signals we have chosen the moments $\langle x \rangle$ and $\langle x^2 \rangle$, corresponding to the first and second position moments of the condensate, as well as the moment $\langle |\psi_t|^2 \rangle$, which we have dubbed ``pointiness''.

In a real system feedback is likely to be limited due to finite detection speed and the ability to dynamically modify the potentials. As with all oscillatory systems controlled with feedback, when the response time of the feedback becomes a significant fraction of the smallest timescale in the dynamics of the system, the control may operate as positive feedback. For this reason, it is only safe to use control system where the dynamics of the relevant fluctuating moments are within the bandwidth of the feedback. For most BEC systems this is not a problem as a control bandwidth of kiloHertz should be sufficient to respond to fluctuations in the system.  The key difficulty in applying feedback in an experimental situation will be the destructive effects of the continuous detection.  It has recently been demonstrated that optical detection cannot be arbitrarily non-destructive, and that for a given atomic spontaneous emission rate, increasing sensitivity over standard techniques requires multi-pass interferometry \cite{hopeET2004,hopeET2005}. 

We solved equations (\ref{eqpsitrapped})--(\ref{eqreservoir}) both with and without feedback terms using a pseudospectral method with a fourth-order Runge-Kutta time step \cite{RungeKutta} with the XMDS numerical package \cite{XMDS} using the atomic properties and loss rates for $^{85}$Rb near a Feshbach resonance, where the interatomic interaction can be tuned with magnetic fields. We used the following physically reasonable parameters for all calculations: $\gamma_t^{(1)} = 7.0\times 10^{-3}$s$^{-1}$, $\gamma_t^{(2)} = 1.7\times 10^{-8}$ms$^{-1}$, $\gamma_u^{(1)} = 7.0\times 10^{-3}$s$^{-1}$, $\gamma_u^{(2)} = 3.3\times 10^{-9}$ms$^{-1}$,  $\gamma_{tu}^{(2)} = 8.3\times 10^{-9}$ms$^{-1}$, $\gamma_p = 5.0$s$^{-1}$, $\kappa_{\mathrm{out}} = 300$, $\kappa_0 = 4.2\times 10^{-4}$ms$^{-1}$, $\lambda=0.01$ms$^{-1}$, $k = 1.0\times 10^{-6}$m$^{-1}$, with a harmonic trapping potential of the form $V_t = m\omega^2 x^2 /2$ where $\omega=50$ rad s$^{-1}$ and $m=1.4095\times 10^{-25}$kg. The simulation region was $2.7\times 10^{-4}$m in length. As the interatomic interaction strengths between different Zeeman levels are unknown for $^{85}$Rb we assume $U_{tt} = U_{uu} = 2U_{tu}$.

\section{Stability analysis without feedback}

First we will consider an analysis where feedback is absent.
Previous analysis has shown that the general rule of stability for pumped atom lasers is that they are unstable when the interatomic nonlinear interaction strength is below a certain critical value, and become more stable as the nonlinear interaction strength increases \cite{Haine2002}. Furthermore, the onset of stability occurs earlier (i.e. at a lower nonlinear interaction strength) if the condensate is preferentially pumped towards the center of the trap; the more narrowly defined the pumping region in space, the more stable is the laser \cite{Haine2003}.

\begin{figure}
\begin{center}
\includegraphics[width=8cm,height=12cm]{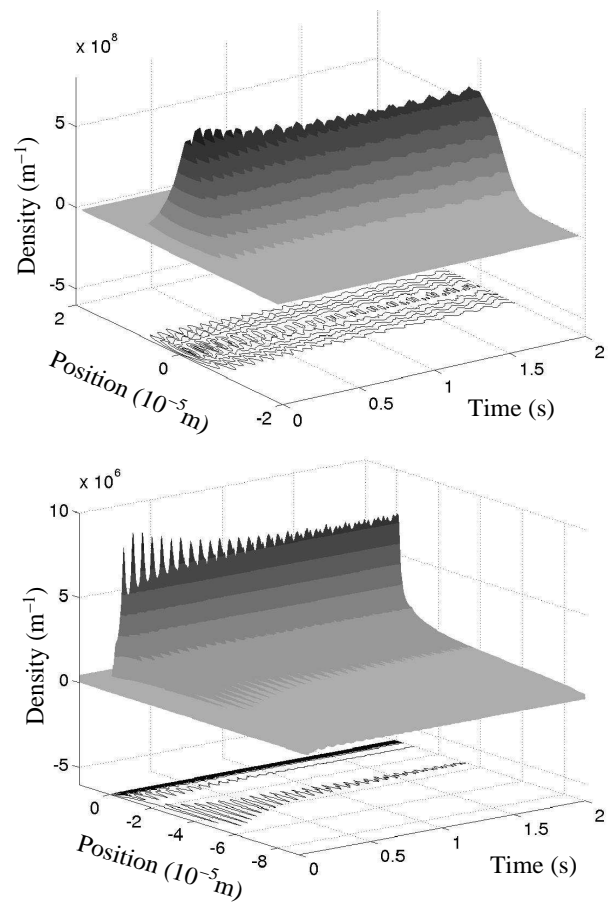}
\caption{General behaviour of the condensate and outcoupled fields. Shown are the condensate density over a two second period (top) and the outcoupled field density over the same period (bottom). Note that the fluctuations of the untrapped field follow those of the trapped field. System parameters were $a=1.0\times 10^{-10}$, $\sigma=9.0\times 10^{-6}$m, $r=3.7\times 10^8$s$^{-1}$.} 
\label{fig_unstable_stable_crossover_3d}
\end{center}
\end{figure}

Our more detailed analysis broadly confirms these findings.
The general behaviour of the condensate is as follows: the trap rapidly fills with atoms, reaching a certain population governed by a balance between pumping and loss. The population and condensate density are initially fluctuating, and these fluctuations will either grow in magnitude or become damped out over time, depending on whether the condensate exhibits long-term stability or not. An example of a borderline stable case is displayed in Figure~\ref{fig_unstable_stable_crossover_3d}, which shows the dynamics of the trapped (condensate) and untrapped (outcoupled) field densities over a period of two seconds. The condensate population and density rise rapidly, and then fluctuate about a steady state value, and then slowly stabilize. The density of the outcoupled field is highest near the outcoupling region, and decreases monotonically in the negative $z$ direction, reflecting the fact that the atoms are accelerating under the influence of gravity. The density fluctuations of the outcoupled field mimic those of the condensate.

\begin{figure}
\begin{center}
\includegraphics[width=8cm,height=7cm]{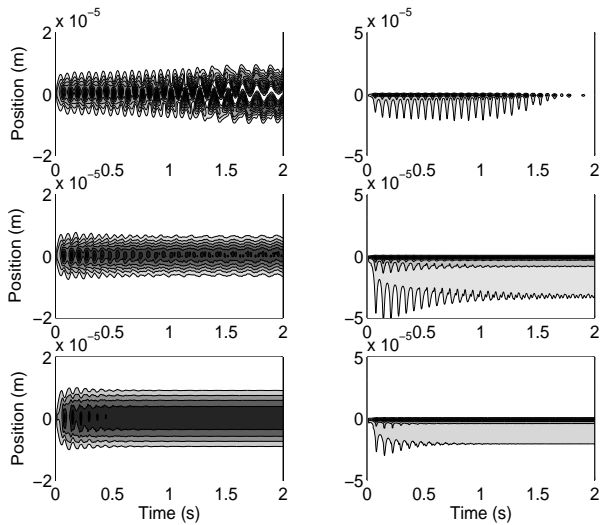}
\caption{Increasing the nonlinear interaction leads to stability. Left, trapped field density; right, untrapped field density. Darker regions indicate higher densities. From top to bottom the scattering lengths are $a=0$, $a=1.0\times 10^{-10}$m, $a=4.65\times 10^{-10}$m. Other parameters were $\sigma=9.0\times 10^{-6}$m, $r=3.7\times 10^8$s$^{-1}$.} 
\label{fig_unstable_stable_crossover}
\end{center}
\end{figure}

The dependence of stability on the interaction strength is also clear in our simulations, as demonstrated in Figure~\ref{fig_unstable_stable_crossover}. The plots show increasing stability as the nonlinear interaction strength between the atoms is increased. The fact that the dynamics of the untrapped field closely follows the fluctuations of the trapped field is again clear. Consequently we need only examine the stability of the condensate in order to determine the stability of the atom beam.

Determining absolute stability from our simulations can be difficult, particularly in borderline cases where long timescales are required in order to determine the asymptotic behaviour of the system. A good technique for determining stability is to decompose the density fluctuations of the condensate into their Fourier components and performing a modal analysis. The ground state of the condensate has the minimum energy allowed and is stable in the trap. As the energy of the condensate is increased, it acquires components of higher order modes, for example breathing modes or sloshing modes. Each mode will experience and gain and loss at a different rate, so between absolute stability and absolute instability exists a regime where the energy of only a subset of all available modes is increasing. This regime is ultimately unstable, but can provide a useful distinction.

\begin{figure}
\begin{center}
\includegraphics[width=8cm,height=8cm]{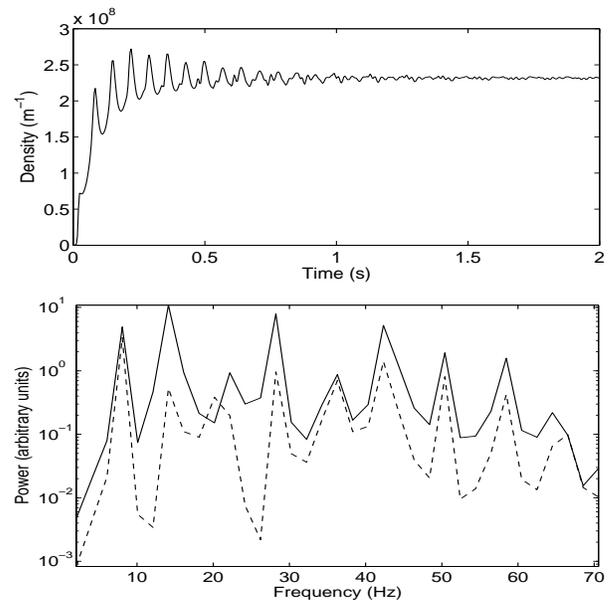}
\caption{Central density of the condensate over a two second period (top), and the associated frequency power spectrum over the period 1.0s--1.5s (bottom left) and 1.5s--2.0s (bottom right).} 
\label{fig_fourier_mode_comparison}
\end{center}
\end{figure}

Figure~\ref{fig_fourier_mode_comparison} shows an example of the Fourier technique. The quantity of interest is the density of the condensate at the center of the trap. A time series of this quantity over a two second simulation is taken, and a Fourier decomposition is made of the two periods 1.0s--1.5s and 1.5s--2.0s. Comparison of these two results can determine which modes are gaining energy and which are losing energy. In the case of Figure~\ref{fig_fourier_mode_comparison}, the energy in every mode has decreased in the second time series, indicating that for this set of parameters the condensate is stable and is attracted to the ground state. 

\begin{figure}
\begin{center}
\includegraphics[width=8.5cm,height=7.5cm]{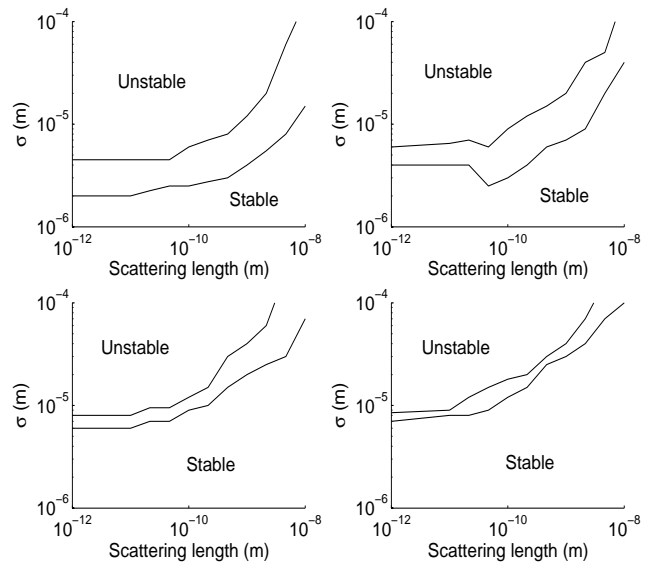}
\caption{The effect of the pumping rate on the stability of the laser when feedback is not present. Above the upper line in each plot the laser is absolutely unstable; below the lower line it is absolutely stable. Between the lines only certain modes are stable. Shown are pumping rates of $r=3.0\times 10^7$s$^{-1}$ (top left), $r=1.0\times 10^8$s$^{-1}$m (top right), $r=6.0\times 10^8$s$^{-1}$ (bottom left) and $r=3.0\times 10^9$s$^{-1}$ (bottom right).} \label{fig_stable_unstable_regions_no_fb}
\end{center}
\end{figure}

Our previous analysis neglected the coupling between the trapped field and the untrapped field on the grounds that the effect of this coupling on the dynamics was usually negligible.  In this approximation the dynamical equations have even parity, and consequently only even modes can be excited.  The odd parity of the gravitational potential means that odd excitations can be excited when the coupling is included.  The odd modes can have a higher gain to loss ratio than the even modes, so including the outcoupling can affect the details of the behaviour of the model.  This is particularly evident near the border of stable and unstable behaviour, where the odd modes can grow over time resulting in instability, where a model without outcoupling would predict stability.  The precise boundaries depend sensitively on the interaction of many variables, including the pumping rate.  As we saw in the absence of spatially selective pumping \cite{Haine2002}, our simulations show that higher pumping rates lead to greater stability. A stability phase diagram is shown in Figure \ref{fig_stable_unstable_regions_no_fb}, showing how the boundaries of stability and instability are affected by altering the pumping rate.  While the long term behaviour becomes asymptotically difficult to integrate near the stability boundaries, the resolution in our simulations is greater than the size of the features in those boundaries.  This means that the stability boundaries are not smooth, and are very sensitive to the details of the system.

\begin{figure}
\begin{center}
\includegraphics[width=8cm,height=8cm]{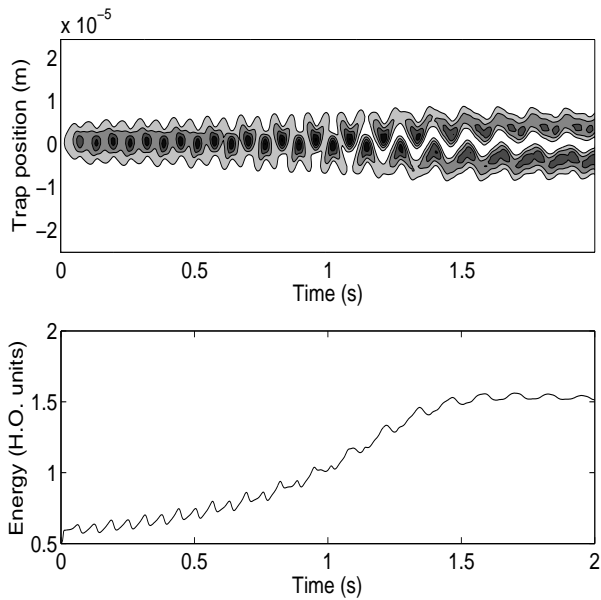}
\caption{An example of a set of parameters where the attractor for the condensate is not the ground state, but rather the first excited state with energy $3/2\hbar \omega$ per particle. The top figure shows the density distribution of the condensate over time, and the lower figure shows the energy per particle in the condensate. Parameters are $a=0$, $\sigma=10.0\times 10^6$, and $r=3.7\times 10^8$.} \label{fig_higher_mode_stability}
\end{center}
\end{figure}

There is also the interesting possibility that there exist stable condensate configurations that are not the ground state. For example, Figure~\ref{fig_higher_mode_stability}(a) shows an example where the trapped condensate is attracted to the first excited trap mode. As this represents a case where the scattering length is zero, the energy of the condensate has no nonlinear component and is simply $3/2 \hbar \omega$, the first excited state of the harmonic oscillator.

\section{Stability analysis with feedback}

Stabilising the condensate can also be achieved by removing energy from it via feedback. If enough energy can be removed the condensate will be in the ground state and stable, and single-mode operation of the atom laser will automatically follow. We will treat the feedback as a set of external potentials which we can continuously modify based on continous measurement of various condensate properties. Ideally we wish to show that if given a certain set of controls, it is possible to adjust them in such a way that the energy of the condensate will always be reduced.
As explained in Section \ref{secintroduction}, it is necessary that we can measure these quantities and modify them on a timescale shorter than the shortest relevant timescale of the moments we are using as error signals.

We begin by considering the case where the condensate is isolated, so that there is no pumping, damping or outcoupling present. In this case condensate evolves via
\begin{equation}
i\hbar \frac{d}{dt}\psi({\mathbf r},t) = (\hat{H}_0 + \hat{V}_{fb})\psi({\mathbf r},t)
\end{equation}
where $\hat{V}_{fb}$ describes the external feedback potential. $\hat{H}_0$ describes the evolution of the condensate when feedback is not present and is given by
\begin{equation}
\hat{H}_0 = \hat{T} + V_0({\mathbf{r}}) + U_0|\psi|^2, \hspace{0.5cm}\hat{T} = -\frac{\hbar^2}{2m}\nabla^2
\end{equation}
where $V_0$ is the trap potential and $U_0$ the nonlinear interaction strength. Defining the energy of the system as
\begin{equation}
E_0(\psi) = \langle \hat{T} + V_0 \rangle + \frac{1}{2}U_0\langle |\psi|^2\rangle
\end{equation}
it is possible to show that once feedback is applied, the rate of change of energy is given by \cite{Haine2004}
\begin{equation}
\frac{dE_0}{dt} = -\frac{i\hbar}{2m} \int V_{fb} (\psi^{\ast} \nabla^2 \psi - \psi \nabla^2 \psi^{\ast}) d^3 {\mathbf{r}}. \label{eqdEdt1}
\end{equation}
As mentioned in Section \ref{secintroduction} we choose our feedback potential to be given by
\begin{equation}
V_{fb} = a_1(t) x + a_2(t) x^2 + b(t) |\psi_t|^2.
\end{equation}
If the parameters $a_1(t)$, $a_2(t)$, and $b(t)$ are chosen to be
\begin{eqnarray}
a_1(t) &=& c_1 \left[ \frac{d \langle x \rangle }{dt} \right], \label{eqc1} \\
a_2(t) &=& c_2 \left[ \frac{d \langle x^2 \rangle }{dt} \right], \label{eqc2} \\
b(t) &=& c_3 \left[ \frac{d \langle |\psi_t|^2 \rangle }{dt} \right] \label{eqc3}
\end{eqnarray}
where $c_1$, $c_2$ and $c_3$ are positive constants and $\langle q\rangle = \int \psi_t^{\ast} q \psi_t dx$, then the rate of change of energy becomes
\begin{equation}
\frac{dE_0}{dt} = -c_1\left[ \frac{d \langle x \rangle }{dt} \right]^2 -c_2 \left[ \frac{d \langle x^2 \rangle }{dt} \right]^2 - \frac{c_3}{2} \left[ \frac{d \langle |\psi_t|^2 \rangle }{dt} \right]^2.
\end{equation}
This is clearly non-positive and demonstrates that in this system the energy of the condensate can always be removed with this feedback scheme.

In what follows, we will choose the constants to be
\begin{eqnarray}
c_1 &=& 2\sqrt{m(m\omega^2 +2c_2 d\langle x^2 \rangle/dt)} \label{eqc1specific} \\
c_2 &=& m^2 \omega^2 / \hbar \\
u_1 &=& \hbar^2 / m \omega N,
\end{eqnarray}
where $N$ is the number of particles in the condensate. The values of $c_1$ and $c_2$ are chosen to ensure critical damping of trap oscillations in the absence of any nonlinear interaction between atoms, and $u_1$ is chosen such that it is an experimentally feasible nonlinearity that is efficient in removing energy from a condensate in the absence of pumping and damping. On the occasions where the term proportional to $c_2$ in (\ref{eqc1specific}) is negative and large enough to cause the quantity under the square root to be less than zero, $c_2$ is set to zero.

The previous analysis becomes more complex, however, when pumping and loss terms are introduced. Now when the $dE_0/dt$ is calculated one finds that there are two seperate contributions, one involving the feedback potential and the other involving the pump and loss terms. These two contributions are decoupled; that is there is no term involving a combination of $V_{fb}$ and pump or loss terms. Consequently the two can be considered independently. The first contribution is due to the feedback potential, and in the notation of Equations (\ref{eqpsitrapped}) and (\ref{eqpsiuntrapped}) is given by
\begin{eqnarray}
\frac{d E_0}{dt}|_{fb} &=& \frac{i}{\hbar} \int \Big( V_{fb} (\kappa_{\mathrm{out}} \psi_t^{\ast}\psi_u - \kappa_{\mathrm{out}}^{\ast} \psi_t \psi_u^{\ast}) \nonumber \\
& +& \left. \frac{i\hbar}{2m}  V_{fb} ( \psi_t \frac{d^2}{dx^2} \psi_t^{\ast} - \psi_t^{\ast} \frac{d^2}{dx^2}\psi_t) \right)  d^3 {\mathbf{r}} \label{eqdEdtnoloss}.
\end{eqnarray}
The second part of (\ref{eqdEdtnoloss}) is equal to (\ref{eqdEdt1}), which is known to be always non-positive in our scheme. The first part of (\ref{eqdEdtnoloss}) represents the interaction between the outcoupling and the feedback, and can be positive or negative depending on the wavefunctions of the trapped and untrapped fields.

The second contribution to $dE_0/dt$ is independent of the feedback potential and can be broken into six parts; five parts proportional to the loss terms $\gamma_t^{(1)}$, $\gamma_u^{(1)}$, $\gamma_t^{(2)}$, $\gamma_u^{(2)}$, and $\gamma_{tu}^{(2)}$, and one part proportional to the pumping rate $\kappa(x)n(x)$. The exact form of the terms is lengthy and will not reproduced here. The crucial point is that some of terms can be positive depending on the form of the trapped and untrapped fields. Consequently the feedback cannot be guaranteed to reduce the energy of the condensate in all circumstances, although the smaller the pumping, outcoupling and loss terms, the better it will do.

\begin{figure}
\begin{center}
\includegraphics[width=8cm,height=8cm]{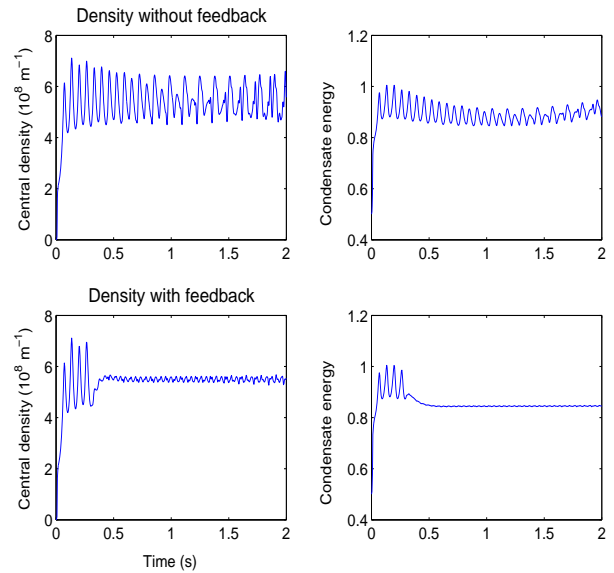}
\caption{The effect of feedback. Top row: central density of the condensate and its energy when no feedback is present. Bottom row: central density of the condensate and its energy when feedback is turned on at $t=0.3$s. Note that ground state energy for this system is 0.839$\hbar \omega$ per particle after the particle number has reached steady state (calculated using an imaginary time algorithm). Parameters used are $r=3.7\times 10^8$s$^{-1}$, $\sigma=9.0\times 10^{-6}$m, $a=4.65\times 10^{-11}$m.} 
\label{fig_feedback_vs_no_feedback}
\end{center}
\end{figure}

\begin{figure}
\begin{center}
\includegraphics[width=7cm,height=6cm]{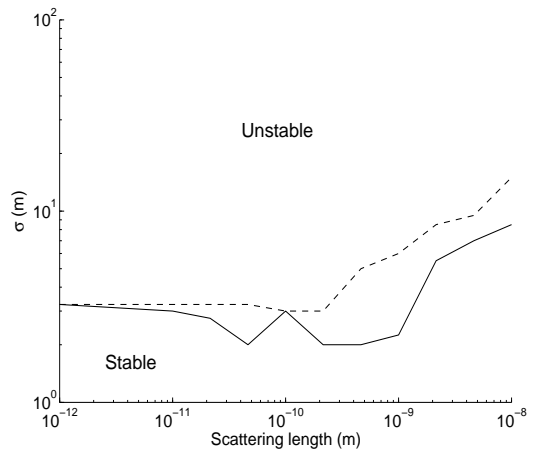}
\caption{How adding feedback improves stability. Dashed line is the boundary of stability with feedback; solid line the boundary of stability without feedback. Pumping rate was $r=6.0\times 10^7$s$^{-1}$.} 
\label{fig_feedback_stability_improvement}
\end{center}
\end{figure}

Although the feedback scheme does not produce stable steady state operation in all parameter regimes, it is still remarkably efficient at stabilising fluctuations in the condensate. Figure~\ref{fig_feedback_vs_no_feedback} shows an example of feedback being applied to a condensate which is in a region of parameter space where it is normally highly unstable.
The introduction of feedback will often damp the condensate fluctuations by up to two orders of magnitude or more, making it useful where all that is required is an ``effective'' stability requirement. That is, if one requires the atom laser to be in single mode for a short period of time, say seconds to minutes, then this feedback scheme can be very useful. If, on the other hand, the required criterion is absolute stability, then this scheme offers a marginal improvement. In general, the introduction of feedback does not greatly alter the stability class of a particular set of parameters. If a condensate is unstable, such that there exist non-ground state modes whose energy increases with time, then applying feedback will drastically damp the oscillations and slow their rate of growth, but given time the condensate will still show the same level of fluctuations as before. Nevertheless, there are regions of parameter space which show an absolute improvement in stability when feedback is applied. Figure~\ref{fig_feedback_stability_improvement} shows an example of such a case.

\section{Conclusion}
We have numerically simulated a pumped atom laser, taking into account the back action of the outcoupled beam as well as a variety of loss mechanisms, in order to determine what factors act to stabilize the laser. As well as considering the effects of altering the pumping envelope, the pump rate, and the atomic scattering length, we introduced a feedback scheme which dynamically alters the trap parameters and atomic scattering lengths in order to remove energy from the condensate and reduce fluctuations.

As noted in previous work, the three significant determiners of stability are the atomic scattering length, the pumping rate and the shape of the pumping envelope.  This does not change significantly in the presence of the feedback scheme.  Stability increases with scattering length and pumping rate, and also increases as the width of the pumping region is decreased. The latter occurs because a narrow pump region preferentially excites lower energy modes at the expense of higher energy modes. In contrast to previous work, we included the effect of the backaction of the outcoupled beam on the condensate, and demonstrated that odd modes are the first to become unstable.

Introducing a feedback scheme had mixed results. It is certainly highly efficient at damping fluctuations in the condensate, leading to approximately single-mode operation. However in many cases it does not change the stability class of a set of parameters. That is, if a particular combination of pump rate, pump envelope and scattering length is known to lead to an unstable condensate, where the fluctuations grow with time, applying feedback will drastically reduce the rate at which the fluctuations grow, but the system is still ultimately unstable.

\end{document}